\documentclass{elsart}

\usepackage{graphics}
\usepackage{epsfig}
\usepackage[dvips]{color}
\usepackage{amssymb}

\def\url#1{{\ttfamily\def\/{/\discretionary{}{}{}}#1}}

\newcommand{\lppr}{\stackrel{<}{\scriptstyle \sim}}
\newcommand{\gppr}{\stackrel{>}{\scriptstyle \sim}}

\begin{document}

\begin{frontmatter}
\title{A possible disk mechanism for the 23d QPO in Mkn~501}
\author{J.H. Fan$^1$, F.M. Rieger$^{2}$, T.X. Hua$^1$, U. C. Joshi$^{1, 3}$, }
\author{J. Li$^1$,
        Y.X. Wang$^{4}$, J.L. Zhou$^{1}$, Y.H. Yuan$^1$, J.B. Su$^1$,
        Y.W. Zhang$^1$}
\thanks{{Corresponding author: J.H. Fan, email: fjh@gzhu.edu.cn}}
\address{1. Center for Astrophysics, Guangzhou University,
 Guangzhou 510400, China\\
2. UCD School of Mathematical Sciences, University College Dublin,
 Belfield, Dublin 4, Ireland\\
3. Physical Research Laboratory, Ahmedabad 380 009, India\\
4. College of Science and Trade, Guangzhou University, Guangzhou
 511442}
\date{Received/Accepted}

\begin{abstract}
  Optically thin two-temperature accretion flows may be thermally and
  viscously stable, but acoustically unstable. Here we propose that
  the O-mode instability of a cooling-dominated optically thin
  two-temperature inner disk may explain the 23-day quasi-periodic
  oscillation (QPO) period observed in the TeV and X-ray light curves
  of Mkn~501 during its 1997 high state. In our model the relativistic
  jet electrons Compton upscatter the disk soft X-ray photons to TeV
  energies, so that the instability-driven X-ray periodicity will lead
  to a corresponding quasi-periodicity in the TeV light curve and
  produce correlated variability. We analyse the dependence of the
  instability-driven quasi-periodicity on the mass (M) of the central
  black hole, the accretion rate ($\rm{\dot{M}}$) and the viscous
  parameter ($\alpha$) of the inner disk. We show that in the case of
  Mkn~501 the first two parameters are constrained by various
  observational results, so that for the instability occurring within
  a two-temperature disk where $\alpha=0.05-1.0$, the quasi-period is
  expected to lie within the range of 8 to 100 days, as indeed the
  case. In particular, for the observed 23-day QPO period our model
  implies a viscosity coefficient $\alpha \leq 0.28$, a sub-Eddington
  accretion rate $\dot{M} \simeq 0.02\,\dot{M}_{\rm Edd}$ and a
  transition radius to the outer standard disk of $r_0 \sim 60 \,r_g$,
  and predicts a period variation $\delta P/P \sim 0.23$ due to the
  motion of the instability region.
\end{abstract}

\end{frontmatter}

\section{Introduction}
 Strong variability is one of the common observational properties of
 TeV emitting Active Galactic Nuclei (AGNs) \cite{cat99}.
 In many cases, the highly variable high-energy { gamma-rays and
 the X-rays appear to be correlated with no time delays evident on
 day-scales}, suggesting that the $\gamma$-rays { may} result
 from inverse Compton upscattering of lower energy photons.
{ The first TeV blazar, for example, that was observed
simultaneously in multiple bands from radio to TeV gamma-rays is
Mrk~421. The first campaign, conducted in 1994 \cite{mac95} on
Mrk~421, revealed correlated variability between the keV X-ray and
TeV gamma-ray emission. The gamma-ray flux varied by an order of
magnitude on a timescale of 2 days and the X-ray flux was observed
to double in 12 hr. On the other hand, the high-energy gamma-ray
flux observed by EGRET, as well as the radio and UV fluxes, showed
less variability than the keV or TeV bands. Another multiwavelength
campaign on Mrk~421 performed in 1995 revealed another coincident
keV/TeV flare \cite{buk96,taka96}. The UV and optical bands also
showed correlation during the flares. With the detection of TeV
emission from Mrk~501 \cite{qui96}, several multiband campaigns were
organized on Mrk~501 to verify whether such a behavior is a general
property of TeV-emitting blazars or whether it is unique to Mrk
421.}


 The gamma-ray blazar Mkn~501, detected as a strong TeV emitter in
 1995 \cite{qui96}, is one of the closest ($z = 0.0337$) and
 brightest BL Lacertae objects. Historically (i.e., prior to 1997),
 its spectral energy distribution (SED) $\nu\,F_{\rm \nu}$ resembles
 that of X-ray selected BL Lac objects, having a peak in the extreme
 UV--soft X-ray energy band \cite{kat99,sam96}. { Earlier} optical
 observations of Mkn~501 have shown variations of up to $1.^{m}3$
 and polarized emission up to $P_{\rm opt} \simeq 7 \%$ \cite{fan99}.

 During its active state in 1997, where Mkn~501 was monitored in the
 2-10 keV X-ray band by the all sky monitor(ASM) on board RXTE and
 in the TeV energy band by several Cherenkov telescope groups
 \cite{aha99a,cat97,dja99,kraw00,pro97,rxte99}, both X-rays and TeV
 gamma-rays increased by more than 1 order of magnitude from
 quiescent level \cite{cat97,pia98}. Analysis of the X-ray and TeV
 data showed, that the variations were strongly correlated between
 both bands, { yet only marginally correlated with the optical UV
 band} \cite{cat97,dja99,kraw00,pet00}.
{ While the synchrotron emission peaked below 0.1 keV in the
quiescent
 state, in 1997 it peaked at ~100 keV. This is the largest shift ever
 observed for a blazar \cite{pia98}. Earlier investigations of the
 1997 April flare in Mkn~501 \cite{pia98}, showed that its origin may
 be related to a variation of $\gamma_{max}$ together with an increased
 luminosity and a flattening of the injected electron distribution.}


During the 1997 high state, the X-ray and TeV light curves displayed
a quasi-periodic signature \cite{hay98,kra99,pro97}. A detailed
periodicity analysis, based on the TeV measurements obtained with
all Cherenkov Telescopes of the HEGRA-Collaboration and the X-ray
data with RXTE, was performed by Kranich et al. \cite{kra99,kra01}
and more recently also by Osone \cite{oso06} including Utah TA data.
The results indeed strongly suggest the existence of a 23 day
periodicity, with a combined probability of $p \simeq 3 \times
10^{-4}$ in both the TeV and the X-ray light curves covering the
same observational period \cite{kra99,kra01}. \footnote{{ Note that
no QPOs have been seen by MAGIC during 1998-2000, when the source
was not very active \cite{alb07}, suggesting that the QPOs only
occur during a very active stage.}} { Rieger and Mannheim ~(2000)
have shown that the origin of these QPOs may be related to the
presence of a binary black hole in the center of Mkn~501. While this
may well be possible, we explore here an alternative scenario, where
the observed QPOs are related to accretion disk instabilities.}

 In a seminal paper, Shapiro et al.~(1976) (SLE) have pointed
 out that a hot (Compton cooling-dominated) optically thin
 two-temperature accretion disk may be present in the inner region
 of the standard optically thick disk \cite{sha73}, whenever the SLE
 inequality $\alpha^{1/4} \dot{M}_{*}^{2}M_{*}^{-7/4} \geq 0.6$ is
 satisfied, where $\dot{M}_{*} = \dot{M}/(10^{17} \rm{g} \rm{s}^{-1})$,
 $M_{*}=M/(3 M_{\odot})$, $\dot{M}$ is the accretion rate of
 the disk, $M$ the mass of the central black hole, and $\alpha$ the
 viscosity parameter, constrained by the model to lie within the
 range $ 0.05 < \alpha < 1$. Later work \cite{pir78,pri76} has
 shown that the SLE configuration might be thermally unstable
 (although less than the standard disk) if the simple standard
 viscosity prescription is employed. However, relatively small
 changes in the viscosity law can already ensure a stable configuration
 \cite{pir78}, in particular, if stabilizing effects of a disk wind
 are fully taken into account.
 A kind of SLE two-temperature disk structure may thus well exist in
 the inner region of BL Lac type objects \cite{wan91,wanu91,cao03}
 and provide a possible explanation for the X-ray and TeV variability
 phenomenon in Mkn~501. For, firstly, Compton processes in a inner
 two-temperature disk with electron temperature $T_e$ of about $10^{9}$
 K (and ion temperature one or two orders of magnitude higher) can
 produce (steep) X-ray power-law spectra, in contrast to the standard
 disk model that can only produce emission up to the optical-UV band
 \cite{sha76,wan91}.
 Secondly, analysis of the linear stability of an optically thin
 two-temperature disk around a compact object shows that the disk
 is subject to a radial pulsational instability (inertial-acoustic
 mode instability) \cite{wu97}. This possible mode of pulsational
 overstability, in which radial disk oscillations with local
 Keplerian frequencies become unstable against axisymmetric
 perturbations, occurs if the viscosity coefficient increases
 sufficiently upon compression \cite{kat78}. In this case,
 thermal energy generation due to viscous dissipation increases
 during compression, leading to amplification of the
 oscillations analogous to the role played by nuclear energy
 generation in stellar pulsations. As we demonstrate below, the
 occurrence of such a type of disk oscillation may well account for
 some quasi-periodic time variability in AGNs in a way similar to
 Galactic black hole candidates \cite{blu84,hon92,man96,yan97}.

\section{Model and Results}
\subsection{Model}
Due to the Lightman \& Eardley secular instability \cite{lig74,lige74}
of the standard geometrically thin, optically thick disk model
\cite{sha73}, a two-temperature disk configuration is likely to be
present in the inner region of the standard disk. As shown by
Shapiro et al.~\cite{sha76}, the outer boundary of such a two-temperature
disk is determined by
\begin{equation}
r_{0*}^{21/8}\zeta^{-2}(r_0)\approx
3.85 \times 10^6\alpha^{1/4}M_{7}^{-7/4}\dot{M}_{24}^2.
\end{equation}
where $r_{0} = r_{0*}{\frac{GM}{c^{2}}}$ is the radius of the outer
boundary, $M_{7}$ is the mass of the central black hole in units of
$10^{7}M_{\odot}$, $\dot{M}_{24}$ is the accretion rate in units of
$10^{24}$ g s$^{-1} = 0.016 M_{\odot}/$yr, $\alpha$ is the viscosity
parameter constrained to be within the range of 0.05 to 1.0, and
$\zeta=1-(\frac{r_{0*}}{6})^{-1/2}$ expresses the boundary condition
that the viscous stress must vanish at the inner edge of the disk.

The total luminosity of a two-temperature disk, integrated from
the inner edge to $r_0$, is of order
\begin{equation}
L_{r0,44}~ =~
{\frac{3}{4}}[1+2(\frac{1}{6}r_{0*})^{-3/2}-18/r_{0*}]\dot{M}_{24}
\end{equation}
where $L_{r0,44}$ is in units of $10^{44}$ erg s$^{-1}$. Combining
relations (1) and (2), one obtains
\begin{eqnarray}
\left[1+2(\frac{r_{0*}}{6})^{-3/2}-\frac{18}{r_{0*}}
\right]r_{0*}^{21/16}\left[1-(\frac{r_{0*}}{6})^{-1/2}\right]^{-1}
=2.62 \times 10^{3} L_{r0,44}\lambda
\end{eqnarray}
where $\lambda\equiv(\alpha M_7^{-7})^{1/8}$. Obviously, since $\lambda$
is relatively insensitive to $\alpha$, the outer radius $r_{0*}$ and the
accretion rate $\dot{M}$ can be uniquely determined, once the central
black mass $M$ and the observed radiative output $L_{r0,44}$ are known.

Following Shapiro et al.~\cite{sha76} (see also \cite{wan91}) we
assume, that a hot (cooling-dominated) optically thin two-temperature
region forms within the inner region of the standard disk. Hence, the
classical outer and middle regions of the standard Shakura \& Sunyaev
(SS) disk model \cite{sha73} describe the outer portions of our disk
model. The inner region of the cool SS disk, whose outer boundary lies
at the point where the radiation pressure $P_R$ matches the gas pressure
$P_G$, extends inwards to the radius $r_0$, where $P_R=3 P_G$. The radius
$r_0$ marks the outer boundary of the two-temperature inner region, which
then extends inward to the innermost radius $3r_g$ for a non-rotating
black hole. In short, the assumed disk configuration is assumed to be
identical to the standard SS model for $r\ge r_0$, but described by the
two-temperature structure equations for $r<r_0$. The disk thickness, the
density, and the ratio $\frac{h}{r}$ for the equilibrium, two-temperature
inner disk then satisfy the following equations:
\begin{eqnarray}
 h &=& 5.29\times 10^{11} M_{7}^{7/12}\dot{M}_{24}^{5/12}\alpha^{-7/12}
 \zeta ^{5/12}r_*^{7/8}~~\rm{cm}~,\\
 \rho &=& 1.14\times 10^{-11}M_{7}^{-3/4}\dot{M}_{24}^{-1/4}\alpha^{3/4}
 \zeta ^{-1/4}r_*^{-9/8}~~\rm{g}~\rm{cm}^{-3}~\\
 h/r & =
 & 0.316\times\rm{M_{7}^{-5/12}}~\dot{\rm{M}}_{24}^{5/12}~\alpha^{-7/12}~
 \zeta^{5/12}~\rm{r_*^{-1/8}}\,.
\end{eqnarray}

Typically, the emission spectrum of a hot ($T_e \sim 10^9$ K) inner
disk is bremstrahlung unless a soft photon source is present, in
which case the spectrum may become a power-law. In our hybrid
model a fraction of the optical-UV soft photons from the outer SS
disk is assumed to be intercepted by the hot SLE inner disk and
Comptonized to form the X-ray power-law spectrum, similar as in
\cite{sha76,wan91}. If Comptonization is not saturated, as possible
in the presence of a copious soft photon source (such as a cold SS
component), the soft photons will be upscattered into a power-law
distribution over some energy range. The resulting disk spectrum
above the input soft energy $E_s=h\,\nu_s$ is approximately of the
form
\begin{equation}\label{spectrum}
 F_{\rm \nu} \propto \nu^{-\alpha_x} \exp(-h\nu/kT_e)\,,
\end{equation} where $\alpha_x = (2.25+4/[y\,(1+f_T)])^{1/2}-3/2 >0$,
$f_T=2.5\,\theta+1.875\,\theta^2(1-\theta)$, $\theta \equiv kT_e/m_e
c^2$ and $y \sim 1$ for unsaturated Compton cooling, i.e., the spectrum
resembles a power-law with index $\alpha_x$ for $E > E_S$ and $E \ll
k\,T_e$, and shows an exponential cut-off for $E \gppr kT_e$
\cite{sha76,wan91}.
The corresponding SLE disk SED $\nu F_{\rm \nu}$ peaks at $E_p =
(1-\alpha_x)\,k T_e$, i.e., usually at around some tens of keV, with
$E_p$ being higher for smaller indices. In particular, for $kT_e=200$
keV ($T_e=2.3 \times 10^9$ K) one may have $\alpha_x \simeq (2.25+
1.85/y)-1.5 \;\in\; [0.4,0.8]$ for $y \;\in \;[0.6,1.3]$.

As first shown by Kato~(1978), accretion disks can undergo
(radial-azimuthal) pulsational instabilities and thereby cause
quasi-periodic variability. If a fluid element in the disk is
perturbed from its equilibrium position, it oscillates in the
radial direction with the epicyclic frequency, as the difference
between gravitational and inertial (Coriolis and centrifugal)
force appears as a restoring force. Due to this property,
axisymmetric perturbations can propagate in the disk as inertial
acoustic waves. As shown by Kato, a viscous disk can be
pulsationally unstable against this type of oscillations under
certain conditions: By studying the stability properties of an
optically thin disk to axisymmetric, local (wavelength $\lambda
\ll$ radial size of disk) and nearly radial ($v_r \gg v_z$)
oscillations, he found that the disk becomes unstable if the
coefficient of the viscosity increases sufficiently rapidly with
density and temperature, and that the frequency of oscillations
essentially corresponds to the angular frequency $\Omega$ of the
disk, independent of wavelength. More recently, Wu~(1997) has
studied in detail the radial pulsation instability of optically
thin two-temperature accretion disks, demonstrating that in a
geometrically thin, cooling-dominated two-temperature disk the
acoustic O-mode (outward-propagating) is always unstable. It
has been argued, that pulsational overstability might explain
some of the QPO phenomena observed in cataclysmic variables (CVs)
and Galactic black hole systems \cite{blu84,che95,man96,mil97},
and be important for a proper understanding of periodic variability
in AGNs \cite{hon92}. As we show in the the next section, Mkn~501
may indeed represent a promising AGN candidate, where pulsational
overstability becomes apparent.

Numerical simulations of pulsational instability of geometrically
thin, optically thick (one-temperature) $\alpha$ disks show
that inertial acoustic waves are excited and propagate both inwards
and outwards periodically, immediately growing to shock waves and
resulting in a time-varying local accretion rate exceeding the
imposed input value $\dot{M}$ by one or two orders of magnitude
\cite{hon92,mil97}.
Global nonlinear calculations of such systems indeed demonstrate
that (i) the net mass flow changes sign by a significant amount
according to the direction of the oscillatory flow, although the
disk seems to maintain its ability to transport on average the
initially imposed $\dot{M}$ \cite{kle93}, and that (ii) the
oscillations also induce small relative changes in other variables
(e.g., temperature or surface density) \cite{mil97}. In particular,
the luminosity variations caused by overstable oscillations in such
systems are expected to be (only) at the percent level, even if
local variables such a mass accretion rate and radial velocity
change significantly \cite{che95,kle93,mil97}. Unfortunately, to
our knowledge, no simulation of pulsational overstability for a
SLE type, optically thin, cooling-dominated two-temperature disk
configuration has been performed up to now. This makes it difficult
to draw solid (quantitative) conclusions for the SLE case as the
simulations performed so far consider physical environments quite
different from that of a hot, optically thin, gas-pressure supported
two-temperature accretion disk. Global simulations of optically thin,
two-temperature disk configurations are certainly required to remedy
this problem. Nevertheless, it seems very likely that significant
changes in local accretion rate during an oscillatory period represent
a genuine qualitative feature associated with pulsational overstability.
If so, then substantial quasi-periodic changes in the total disk
luminosity might be expected for the SLE case (cf. \cite{sha76},
eqs.[1] and [21]).

In principle, the observable period of the pulsational instability
in a viscous disk is of order \cite{yan97}
\begin{eqnarray}
P & \approx & (1+z)~\alpha^{-1}\Omega^{-1}\nonumber\\
  & = & 5.78 \times 10^{-4}~(1+z)~M_7\,r_{*}^{3/2} \alpha^{-1}
      [\rm {days}]\,,
\end{eqnarray} where $\Omega^{-1}~=~\sqrt{r^3/GM}$ is the local
Keplerian timescale, $r_{*}$, in units of ${\frac{GM}{c^{2}}}$,
is the radius at which the pulsation instability occurs, and $z$
is the redshift of the source. Obviously, since the disk structure
depends significantly on the radius, the period of the pulsational
instability is different from one radius to another. Yet, by using
eq.(3) we can calculate the radius $r_0$, at which the disk becomes
unstable to radial pulsation for a given luminosity, and so derive
an upper limit on the range of possible periods. In fact, these
upper limits may offer a very useful guide, since numerical disk
simulations suggest that the oscillations may always be trapped
near the boundaries of the discs \cite{kle93,pap86}.

\subsection{Application to Mkn 501}

 The $\gamma$-ray blazar Mkn 501 was the second source after Mkn 421
 to be detected at TeV energies in 1995 by the Whipple observatory
 \cite{qui96}. In 1997 the source came into much attention when it
 went into a remarkable state of strong and continuous flaring activity,
 becoming the brightest source in the sky at TeV energies
 \cite{aha99a,aha99b,cat97,dja99,hay98,pro97}.
 BeppoSAX observations during a strong outburst in April 1997 showed,
 that the spectrum was exceptionally hard ($\alpha\leq 1$, $F_{\nu}
 \propto \nu^{-\alpha}$) over the range 0.1-200 keV, indicating that
 the X-ray power output peaked at $100$ keV or higher energies
 \cite{pia98}.
 These observations implied that the peak frequency has moved up by
 more than two orders of magnitude (persistent over a timescale of
 $\sim 10$ d) and that the (apparent) bolometric luminosity of
 Mkn~501 has increased by a factor of $\gppr 20$ compared to previous
 epochs \cite{kat99,kraw00,pia98,sam96,tav01}. Optical observations
 on the other hand indicated, that the source was still relatively
 normal and thus suggested that the variations are confined to
 energies above $\sim 0.1$ keV \cite{pet00}. Further observations
 with BeppoSAX in April-May 1998 and May 1999 during periods of
 lower TeV flux showed that the peak frequency has decreased to
 $\sim 20$ and $\sim 0.5$ keV, respectively \cite{tav01}.

 A comprehensive analysis of the temporal characteristics of the
 gamma-ray emission from Mkn~501 in 1997 showed that the TeV emission
 varied significantly on time scales of between (5-15) hours
 \cite{aha99a}. If this short time variability is via inverse Compton
 processes related to accretion disk phenomena (see \S~3), we can
 estimate an upper limit for the black hole mass in Mkn~501 by taking
 the variability timescale to be of order of the Keplerian orbital
 period at the innermost stable orbit $r_{\rm ms}$, suggesting
 \begin{equation}\label{mass}
  M_{7} \leq 7.9 \left(\frac{\Delta t}{10\,\rm{hr}}
                        \right) \frac{1}{(1+z)}
 \end{equation} for a non-rotating black hole where $r_{\rm ms}=3\,
 r_{g} = \frac{6GM}{c^2}$, and thus a black hole mass of $M \leq
 (3.8 -11.5) \times 10^{7} M_{\odot}$ for the (5-15) hour timescale
 observed. Note, that this mass upper limit may be somewhat higher,
 if a pseudo-Newtonian potential is employed, in which case one
 may find $M \leq (5.7-17.2) \times 10^7 M_{\odot}$. {
 Eq.~(\ref{mass}) essentially assumes that the observed short-time
 variability is dominated by processes occurring close to the
 innermost stable orbit. This seems justified as long as the
 associated timescale ($t \sim 3.6 \times 10^4$ sec in the lab.
 frame) is larger than (i) the transverse light crossing time
 of the source as measured in the lab. frame, i.e., $t_l \sim r_b/
 [\Gamma_b\,c] < t$, where $r_b \sim z/\Gamma_b$ with $\Gamma_b
 \sim 15$ the bulk Lorentz factor and $z \sim 5 \times 10^2 r_g$
 the distance from the central engine, cf. \cite{der94}, and (ii)
 the characteristic shock acceleration timescale $t_{\rm acc}
 \lppr 10^3\,r_{\rm gyr}/(\Gamma_b c) < t_l$ \cite{rie07}, where
 $r_{\rm gyr}$ is the electron gyroradius for intrinsic magnetic
 field strengths $\gppr 0.01$ G and comoving electron Lorentz
 factors of $\sim 10^5$ (see below), which seems indeed to be the
 case. Based on the above noted, as well as related considerations
 in \cite{rie03}}, we adopt a black hole mass of $M = 9 \times
 10^7 M_{\odot}$, { for which $r_g \simeq 2.66 \times 10^{13}$
 cm,} as fiducial value in our calculations below.

 Detailed observations of Mkn~501 during the 1997 active phase
 showed that its X-ray emission was highly variable, although the
 soft X-ray flux (up to a few keV) did not change dramatically
 \cite{kraw00,lam98,mas04,pia98}. BeppoSAX observations in 1997
 give (2-10) keV fluxes in the range $(1.69-5.24)\times 10^{-10}$
 erg cm$^{-2}$ s$^{-1}$, with an average value of $2.50\times
 10^{-10}$ erg cm$^{-2}$ s$^{-1}$ \cite{mas04}, corresponding to
 a mean (bolometric) luminosity of $<L_{X}> = 6.2 \times 10^{44}$
 erg s$^{-1}$, { assuming H$_{0}$ = 72 km/s/Mpc and a flat
 Universe with $\Omega_M =0.27$.}
 If the observed periodicity in the (2-10) keV band is caused by
 a pulsational disk instability, a non-negligible fraction $f<1$
 of this X-ray luminosity has to be produced by the inner disk,
 the other part being provided by the relativistic jet.
 Each surface of the disk then produces a characteristic (2-10)
 keV luminosity of $L_{\rm 2-10\,keV} = 3.1 \times 10^{44} \,f$
 erg s$^{-1}$. Simulations of pulsational instability discussed
 at the end of \S~2.1 indicate that the oscillations are
 accompanied by periodic, large amplitude variations in local
 accretion rate (exceeding the average $\dot{M}$ by over an
 order of magnitude) and possibly trapped near the outer edge
 of the disk. Simple quantitative modelling then suggests that
 the total luminosity of a SLE type disk may vary by up to a
 factor of a some few, i.e., $f$ may be as small as $\sim 0.1$.
 Using the observed soft X-ray power-law index $\alpha_x \simeq
 0.7$ \cite{kraw02,mas04,pet00}, we  may estimate the observationally
 required, total integrated SLE disk luminosity $L_{r0}$ by
 calculating the ratio $\eta \equiv L_{\rm 2-10\,keV}/L_{r0}$ of
 the soft X-ray to total disk luminosity from the emitted disk
 spectrum (see eq.~[\ref{spectrum}]), which for $k T_e \simeq
 200$ keV gives $\eta \simeq 1/5$. This suggests that the total
 (average)  SLE disk luminosity is of order $L_{r0} \simeq 16
 \times 10^{44} \,f$ erg s$^{-1}$.

\begin{figure}
   \centering
   \includegraphics[width=11cm]{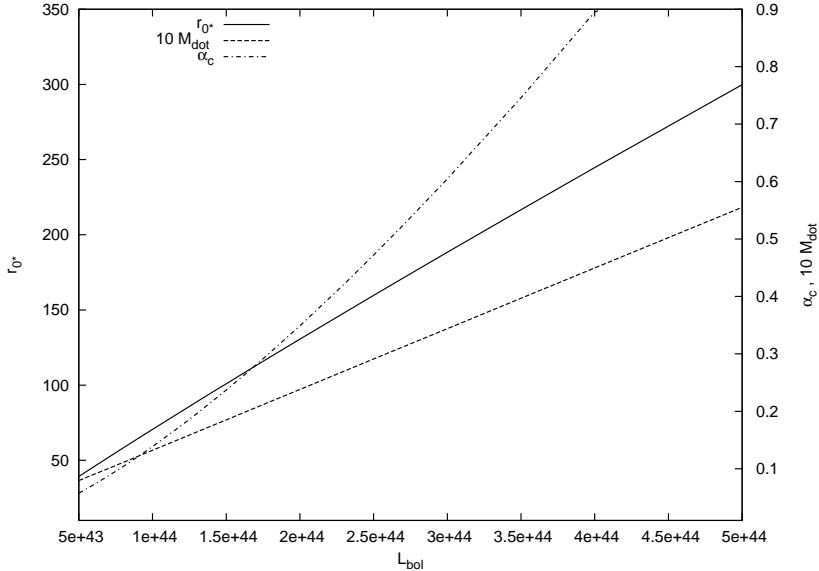}
   \caption{Outer SLE radius $r_{0*}$, accretion rate $M_{\rm dot}
    \equiv \dot{M}/\dot{M}_{\rm Edd}$ and upper limit for the
    viscosity coefficient $\alpha\leq \alpha_c$ as a function of
    the luminosity of a two-temperature SLE disk. Parameters
    $M=9 \times 10^7\,M_{\odot}$ and $P_{\rm obs}=23$ d have been
    employed in the calculations.}
      \label{lplot}
   \end{figure}

 As shown below, low $f$ values are more preferable in the
 hybrid SLE model and we will thus henceforth adopt a total
 SLE disk luminosity $<L_{r0}>= 1.75 \times 10^{44}$ erg  s$^{-1}$
 ($f=0.11$) for explicit calculation (see however also
 Fig.~\ref{lplot} for the more general case).
 Given the constraints on the viscous parameter $\alpha$ and
 using $M = 9.0 \times 10^{7} \rm{M_{\odot}}$, we may then
 obtain the following results for the transition radius
 $\rm{r_{0*}}$, the maximum pulsational period P and the ratio
 $\frac{h}{r}$, namely, for $\alpha = 0.05$ we have: $\rm{r_{0*}}
 =97$, P = 100 days, ${\frac{h}{r_0}}=0.55$; while for $\alpha
 =1.0$ one finds: $\rm{r_{0*}}=130$, P=8 days, ${\frac{h}{r_0}}
 = 0.09$, respectively. In both cases the accretion rate is
 similar, i.e., $\dot{M}_{24}\simeq 2.7$, corresponding to
 $\dot{M} \simeq 0.02 \dot{M}_{\rm Edd}$, and consistent with
 the condition (see \S~1) for a SLE configuraton to exist.
 We note that for BL Lac objects the transition radius between
 an inner optically thin two-temperature and an outer standard
 disk has been estimated to lie within the range $r_{0*} \simeq
 (40-150)$ for $\dot{M}/\dot{M}_{\rm Edd}=0.01$ \cite{cao03},
 which seems well consistent with the values derived above.
 Based on these results, we can also roughly estimate the
 luminosity $L_{ss} \sim 2 \pi r^2 \sigma T^4$ associated with
 the SS disk component at $r\geq r_0$: Employing the thin
 cold disk temperature relation $T \simeq 6.3 \times 10^5\,
 \dot{m}^{1/4}(10^8 M_{\odot}/M)^{1/4}(r_g/r)^{3/4} \simeq 1.3
 \times 10^4$ K with $r =r_0 \simeq 50\,r_g$ one finds that
 $L_{ss} \simeq 1.8 \times 10^{43}$ erg/s, and that the emission
 is maximized at frequencies around $f \simeq 3 \,k\,T/h
 \simeq 8 \times 10^{14}$ Hz. These findings imply a
 Compton energy enhancement factor $A$ considerably larger
 than $L_{r0}/L_{ss} \simeq 10$.

 Spectroscopic data for Mkn~501 suggest, that the normal
 bolometric (photo-ionization-equivalent) luminosity $L_b$
 of the disk component driving the NLR photoionization is of
 order a few times $10^{43}$ erg/s, corresponding to a photon
 number per second that can ionize hydrogen of $N_b(H) \lppr
 L_b/(h\nu_1)\sim 10^{54}$ s$^{-1}$, where $\nu_1=13.6$ eV/$h$.
 Here, $L_b$ has been estimated based on the $L_b$-$L(H\alpha)$
 relation, i.e., $\log L_b = 1.176 \,\log L(H\alpha)-4.91$
 \cite{lao03}, with an observed (mean) narrow H${\alpha}$ line
 luminosity for Mkn~501 of $\simeq 10^{41}$ erg/s \cite{sti93},
 which gives $L_b \sim 2 \times 10^{43}$ erg/s (with substantial
 scatter). We note that this value is very close to the one
 estimated by Barth et al.~(2002) using emission line
 measurements for four nearby FR I radio galaxies. Given the
 properties of the cold SS disk component derived above, the
 main part of the ionizing photons per second is expected to
 come from the SLE component (unless its existence is only
 coupled to active source stages). The number of photons
 emitted by the SLE inner disk per second, which can ionize
 hydrogen is given by $N(H)=\int_{\nu_1}^{\infty} L_{\nu}/[h\,
 \nu] d\nu$, where $L_{\nu}$ is the specific SLE luminosity
 (cf. eq.~[\ref{spectrum}]). Using the values adopted above,
 i.e., $\alpha_x \simeq 0.7$, $k\,T_e \simeq 200$ keV and
 $L_{\rm 2-10 \,keV} = 5.8 \times 10^{44} \rho \,f$ erg s$^{-1}$
 ($\rho \lppr 0.3$ and $\rho = 1$ for the quiescent and
 active source stage, respectively, cf. \cite{mas04}), we
 can find $N(H) \simeq 4.5 \times 10^{55} \rho \,f$ s$^{-1}$,
 which for $\rho \simeq 0.2$ and $f \simeq 0.1$ is comparable
 to $N_b(H)$, and may thus qualify the choice of $f$ above.

 As mentioned in \S~1, detailed periodicity analysis of the
 X-ray and TeV light-curves of Mkn~501 during its high state
 in 1997 has provided strong evidence for a 23-day period in
 both energy bands \cite{kra99,kra01,oso06}. As shown above,
 such a period falls within the range of possible pulsational
 periods, suggesting that it may be well explained by the
 pulsational instability of a two-temperature SLE disk. In
 particular, by using $P=23$ day and eqs.~(3) and (7), we can
 place an upper limit on the allowed viscosity parameter and
 the expected transition radius. For the characteristic values
 derived above, one obtains $\alpha \leq 0.28$, corresponding
 to $\rm{r_{0*}}$ = 115 and ${\frac{h}{r_0}} \sim 0.20$. The
 more general case is shown in Fig.~(\ref{lplot}), where the
 dependency of these results on the total SLE disk luminosity
 has been calculated.

{ The changing SLE disk luminosity will lead to a variable seed
 photon field for Compton upscattering to the high energy regime
 by the nonthermal particles in the relativistic blazar jet. The
 disk radiation entering a moving plasma blob from behind appears
 de-boosted by a factor $\sim \Gamma_b^2$ in the comoving frame of
 the outflowing plasma. It will nevertheless dominate over possible
 (quasi-isotropic) disk photons that are scattered by diffusive gas
 or clouds into the direction of the source, provided the distances
 $z$ to the central engine is smaller than
 \begin{equation}
  z \lppr 7 \times 10^3 \,\left(\frac{15}{\Gamma_b}\right)^2
              \left(\frac{R_{\rm sc}}{1\,\mathrm{pc}}\right)
              \left(\frac{0.001}{\tau_{\rm sc}}\right)^{1/2}
              \;r_g\,,
 \end{equation}
 where $R_{\rm sc}$ is the distance and $\tau_{\rm sc}$ the mean
 scattering depth of the electron scattering cloud \cite{der94}.
 This usally places the emission site at distances between hundred
 (to avoid $\gamma\gamma$ absorption) and some thousand $r_g$.

 As shown by Dermer and Schlickeiser (1993), scattering of X-ray
 ($\gppr 0.1$ keV) photons entering from behind will generally take
 place in both the Thomson and the Klein-Nishina regime for a large
 range of (lab. frame) soft photon energies $\epsilon_{\rm ph}$, i.e.,
 for energies in the range $\Gamma_b m_e c^2/\gamma < \epsilon_{\rm
 ph} < 4 \Gamma_b \gamma m_e c^2$, where $\Gamma_b \sim 15$ is the
 jet bulk Lorentz factor. An full treatment of the scattering process
 is thus a complicated matter and beyond the scope of the paper.
 However, we can still get some order-of-magnitude insight into
 the associated photon luminosity by approximating the scattering
 process by its Thomson limit: The maximum value of the scattered
 photon energy $\epsilon_{\rm ph}^s$ in the lab. frame is given by
 $\epsilon_{\rm ph}^s \simeq 0.5\,\gamma^2\,\epsilon_{\rm ph}$
 \cite{der92}. Upscattering of soft X-ray photons ($\gppr 0.2$ keV)
 to the TeV regime thus requires comoving electron Lorentz factors
 of $\gamma \sim 10^5$. Scattering will always take place in the
 Thomson regime for (lab.frame) soft photon energies $\epsilon_{\rm
 ph} < \Gamma_b m_e c^2/\gamma \lppr 0.08$ keV \cite{der93},
 suggesting that our approximation is indeed not too bad if we
 take an analogous reduction of the Thomson cross-section by a
 factor of two or so into account. For photons entering from
 behind the single inverse Compton power in the comoving blob
 frame for scattering by a relativistic electron with Lorentz
 factor $\gamma$ then becomes, cf. \cite{der94},
 \begin{equation}
  P_{\rm Comp} \sim \frac{c}{6}\,\sigma_{\rm T}\,
                         \frac{\gamma^2}{\Gamma_b^2}\,
                         \frac{L_{\rm r0}}{2 \pi z^2 c}\,.
 \end{equation}
 Suppose, that electrons are continuously accelerated at a shock
 front into a characteristic non-thermal (isotropic) differential
 power law particle distribution in the comoving frame of
 $n(\gamma)=n_0 \gamma^{-2}$ for $10^2 \lppr \gamma_1 \leq \gamma
 \leq \gamma_{\rm max}$ with $\gamma_{\rm max} \sim 2 \times 10^6$
 roughly given by the balance between acceleration and external
 Compton cooling for $B \sim 0.05$ G and $z \sim 500\,r_g$. Due to
 $\gamma_{\rm max} \propto L_{r0}^{-1/2}$, an increase of the SLE
 disk field will lead to a decrease in the maximum Lorentz factor.
 As particle are advected downstream and no longer efficiently
 accelerated, the distribution steepens to $n(\gamma) \propto
 \gamma^{-3}$ for momenta where particles have had sufficiently
 time to cool, resulting into a broken power law electron
 distribution with a break energy of $\gamma_b$. We may estimate
 $\gamma_b$ by setting the (comoving) external Compton cooling time
 scale equal to the (comoving) dynamical time scale $\sim z/
 (\Gamma_b\,c)$, which gives $\gamma_b \sim 2 \times 10^5$. For
 typical values $n_0 \gamma_1 \sim 10^4 \mathrm{cm}^{-3}$, e.g.,
 \cite{pet00,kraw02}, the associated observable Compton VHE
 luminosity can thus be estimated from $L_{\rm obs} \sim \Gamma_b^4
 P_{\rm Comp} N_{\rm eq} \Delta V$ where $N_{\rm b} \sim
 n(\gamma_{\rm b}) \gamma_{\rm b}$ and $\Delta V \sim 4 \pi (z/
 \Gamma_b)^3 \Gamma_b$ is the volume scale of the emitting blob
 measured in the comoving frame. This gives a Compton luminosity
 \begin{equation}
   L_{\rm TeV} \sim L_{\rm r0} \left(\frac{z}{500\,r_g}\right)
               \left(\frac{\gamma_{\rm b}}{10^5}\right)
               \left(\frac{N\,\gamma_1}{10^4 \mathrm{cm}^{-3}}\right)\,,
 \end{equation} which is of the order of the observed VHE luminosity
 above $1.5$ TeV during the 1997 high state \cite{aha99a,pet00}. This
 suggests that changes in the disk photon field driven by a pulsational
 instability may indeed be responsible for the observed modulation of
 the TeV emission. During the quiescent state on the other hand, the
 X-ray flux is typically about ten times smaller (cf. \S 2.2). Hence,
 even if a SLE type disk is present in the quiescent state (see
 however below), the expected TeV contribution due to inverse Compton
 scattering of disk photon is much smaller, perhaps even swamped by
 synchrotron self-Compton emission. The 1997 observations of Mkn~501
 showed that the observed TeV spectrum gradually steepened with
 energy starting at around 3 TeV, approaching a logarithmic spectral
 ($\nu\,F_{\nu}$) slope above 5 TeV of $\simeq 1.7 \pm 0.33 \pm 0.6$
 including systematical and statistical errors \cite{aha99b,kon99}.
 However, as shown for example by Konopelko et al.~(2003), this
 curvature at around 3 TeV is likely to be caused by gamma-ray
 absorption in the intergalactic infrared background field,
 suggesting that the peak in the intrinsic (de-absorped) spectrum
 due to increasing Klein-Nishina effects is around 8 TeV.
 In principle, it appears possible that due to the strong angle
 dependence of the scattered flux on the angle of the impinging
 soft photons, photons from the outer SS disk part, i.e. from disk
 radii $\sim z$, could make an important inverse Compton contribution
 as well. Redoing the calculations presented in \cite{der94} for
 lab.frame angles of the incident photons of $\sim \pi/4$ gives
 a total photon energy density of the soft photons in the comoving
 frame of $u_{\rm ph}^s \simeq (1-\beta_b/\sqrt{2})^2\,u_{\rm ph,
 s}^{\star}$ where $u_{\rm ph,s}^{\star} \simeq L_{\rm SS}(r0/z)/
 (4 \pi z^2 c)$ is the corresponding energy density in the lab.
 frame. As the SLE contribution $u_{\rm ph}^{\rm SLE}$, entering
 from behind, appears de-boosted by a factor of $\simeq 4
 \Gamma_b^2$ \cite{der94} compared to its lab. frame value, the
 ratio of comoving energy densities becomes
 \begin{equation}
   \frac{u_{\rm ph}^s}{u_{\rm ph}^{\rm SLE}} \sim 0.35
           \left(\frac{L_{\rm r0}}{L_{\rm SS}}\right)
           \left(\frac{z}{r_0}\right)\,,
 \end{equation} using $\Gamma_b=15$, $L_{\rm r0}/L_{\rm SS} \sim 10$
 and $z/r_0 \sim 10$, indicating that the SLE part dominates during
 the high state even if we assume that the SS disk is not truncated
 at radii of $\sim 500\,r_g$. Note however, that the latter may well
 be the case if a binary black hole system exists in the center of
 Mkn~501 as argued for by several authors (see discussion).}

\section{Discussion}
 The X-ray to TeV spectra of TeV blazars have been often interpreted
 within a one-zone SSC model, e.g. \cite{blo96,der97,mas97,tav98,tav01}.
 However, if a two-temperature disk structure is present in some of
 these sources, the real situation may be more complex as the X-ray
 radiation from a two-temperature disk, for example, may represent an
 additional, non-negligible source of seed photons for the inverse
 Compton scattering to TeV energies, in particular during active
 source stages, likely to be associated with changes in accretion
 history. Hence our model assumes, that a pulsational instability
 occurring within a two-temperature disk leads to observable, periodic
 variations of its X-ray radiation field. Part of this periodically
 modulated X-ray emission will enter the jet and (in addition to direct
 synchrotron photons) serves as seed photons for Compton-upscattering
 to TeV energies { similar as in \cite{der92,der93}}. Our model thus
 takes it that both, a direct synchrotron self Compton and an external
 Compton contribution are relevant for modelling the SED of Mkn~501,
 the seed photons for external Compton-upscattering consisting of both,
 the infrared-optical seed photons from the (quasi-steady) SS disk
 component and the variable X-ray photons from the two-temperature
 disk component.
 A related, but more simpler scenario, assuming the seed photons for
 inverse Compton scattering to be provided by a (direct) synchrotron
 plus an quasi-steady flux component comparable to the observed
 infrared-optical flux, has been proposed by Pian et al.~(1998) in
 order to account for the different degrees of SED variations of
 Mkn~501 at X-ray and sub-X-ray energies during the April 1997
 outburst (see also \cite{ghi98,kat99}).
 Detailed analysis indeed suggests that one-component SSC models
 cannot fit both the April 1997 SEDs and the lightcurves from X-ray
 to TeV, and that (at least) an additional, moderately variable low
 energy component contributing in the energy range between 3-25 keV
 is required \cite{kraw02,mas04}. A similar conclusion seems to hold
 for the strong X-ray outburst observed in July 1997 \cite{lam98}.
 Interestingly, observations in 1998 also provide evidence for an
 additional component in the optical regime, possible associated
 with the SS disk component in our hybrid disk model { (see
 also \cite{kat01} for an alternative interpretation)}: As shown
 by Massaro et al.~\cite{mas04} optical to X-ray data taken in June
 1998 indicate that the optical spectrum is steep and does not match
 the low energy extrapolation of the X-ray spectrum, hence suggesting
 the presence of different emission components in the optical and
 in the X-ray regime as naturally expected in our model.

 Monitors in both the X-rays and the TeV emission show evidence for
 a 23-day periodicity during the 1997 high state. As demonstrated
 above the 23-day period in the X-ray light curve may be caused
 by a pulsational instability in two-temperature accretion disk,
 and via the inverse Compton process result in the same periodicity
 in the TeV light-curve. A pulsational instability occurring in
 a two-temperature disk with transition radius $r_0 \sim (48-65)\,
 r_{g}$  will result in a recurrence timescale of 8 to 100 days.

 Based on the observed TeV variability we have employed a
 characteristic black hole mass of $9 \times 10^7 M_{\odot}$
 in our calculations. We note that quite different central mass
 estimates for Mkn~501 have been claimed in the literature,
 ranging from several times $10^7$ (mainly based on high energy
 emission properties) up to $10^9\,M_{\odot}$ (based on host
 galaxy observations), see e.g. \cite{cao02,dej99,fal02,fan05,rie03}.
 However, as shown by Rieger \& Mannheim~\cite{rie03} uncertainties
 associated with host galaxy observations may easily lead to an
 overestimate of the central black hole mass in Mkn~501 by a
 factor of three and thus reduce the implied central mass to
 $\simeq (2-3)\times 10^8\,M_{\odot}$, { a value in fact
 recently confirmed by an independent analysis of central mass
 constraints derived from host galaxy observations \cite{woo05}.}
 Moreover, as argued by the same authors some of the apparent
 disagreement in central mass estimates may possibly be resolved
 if a binary black hole system exists in the center of Mkn~501,
 see also \cite{rie00,rie03,vil99}, similar as in the case of
 OJ~287 \cite{sil88}.
 For example, if Mkn~501 harbours a binary system with a more
 massive primary black hole of $\lppr 10^{9} M_{\odot}$ and
 a less massive (jet-emitting) secondary black hole of $\sim
 10^{8}\,M_{\odot}$, the mass ratio $\rho=m/M$ would be of
 order 0.1, which may compare well with the result $\rho<0.25$
 estimated for OJ~287 \cite{liu02}. While our characteristic
 black hole mass employed falls well within the above noted
 range, we note that the SLE pulsational instability model may
 still work successfully, if a higher black hole mass is used.
 For example, if one adopts $M = 3 \times 10^{8} M_{\odot}$ and
 $P=23$ days, one obtains $\alpha \leq 0.07$, $r_{0*}=46$ and
 $\dot{M}\simeq 0.008 \dot{M}_{\rm Edd}$.

 Our analysis is based on a specific disk model (SLE) which is
 open to questions, in particular with respect to its possible
 stability properties. We note however, that a relatively small
 change in the usually employed viscosity description may already
 lead to a thermally stable configuration \cite{pir78}. On the
 other hand, it may as well be possible that the SLE configuration
 represents a quasi-transient phenomenon associated with those
 changes in accretion history that probably initiate the high
 states. An alternative (inner) disk configuration of interest
 may be represented by an optically thin two-temperature ADAF
 solution \cite{nar98,yi99}. Such a configuration can exist for
 accretion rates $\dot{m}= \dot{M}/\dot{M}_{\rm Edd}$ below a
 critical rate $\dot{m}_{\rm crit} \simeq 0.3\,\alpha^2 \simeq
 0.019$, where the canonical ADAF value of $\alpha =0.25$ has
 been employed \cite{nar98}. ADAFs are generally less luminous
 than a standard disk, with the typical ADAF luminosity given by
 $L_A \simeq 0.02\,(\dot{m}/\alpha^2)\,\dot{M}\,c^2$ \cite{yi99}.
 Using the constraints above, the possible ADAF luminosity for
 Mkn~501 becomes $L_A \leq 1.3 \times 10^{43}$ erg/s, which is
 already about an order of magnitude smaller than required by
 the X-ray analysis. This suggests that -- at least during its
 high state -- an optically thin ADAF is not a viable option
 for Mkn~501.

 Based on Eq. (7) we can estimate the variation rate of the
 period due to the motion of the instability
 \begin{eqnarray}
 \delta P/P
  = {\frac{3}{2}} P{\frac{1}{r}} {\frac{\partial{r}}{\partial{t}}}\,.
 \end{eqnarray}
 Using $v_{r} = {\frac{\partial{r}}{\partial{t}}}$, $\dot{M} =
 4 \pi \rho h r v_{r}$ and Eqs. (4) and (5), one finds
 \begin{eqnarray}
 \delta P/P  =0.44
 ~\alpha^{-7/6}~\dot{M_{24}}^{5/6}~M_{7}^{-5/6}~r_{*}^{-1/4}
 \zeta^{-1/6}
 \end{eqnarray}
 For $\alpha$ = 0.28 the relevant parameters result in $\delta P/P=
 0.23$. From the period analysis performed by Kranich et al.~\cite{kra99}
 a $3\sigma$ deviation in period corresponds to 6.67 days. If we take
 this deviation as the intrinsic variation on the periodicity (P),
 then a $\delta P/P$ of ${\frac{6.67}{23}}=0.29$ can be estimated
 from the results by Kranich et al.~\cite{kra99}. As this estimate
 assumes that the deviation of the period is only affected by the
 motion of the instability, while it may in fact be caused by more
 than one effect, our theoretical $\delta P/P$ should not be greater
 than the observational results. We conclude that the observed 23-day
 QPOs in Mkn~501 might be caused by the instability of a two-temperature
 accretion disk. The model presented here may thus offers an
 alternative explanation to the binary-driven helical jet model of
 Rieger \& Mannheim~(2000). Comprehensive computational modelling
 of the pulsational instability in a two-temperature, cooling
 dominated disk will be essential to verify this in more detail.

 Our model predicts that a period correlation in the X-ray and
 $\gamma$-ray should always be present during an active source
 stage, while the period of the QPOs may vary as the instability
 region could change from one high state to the other.

\section*{Acknowledgements}
 We would like to thank Prof. K.S. Cheng,  J.M. Wang, Y. Lu, L.
 Zhang and Dr. D. Kranich for useful discussions, { and the
 anonymous referees for very useful comments that helped to
 improved the presentation.} This work is
 partially supported by the National 973 project (NKBRSF G19990754),
 the National Science Fund for Distinguished Young Scholars (10125313),
 the National Natural Science Foundation of China (10573005, 10633010),
 the Fund for Top Scholars of Guangdong Province (Q02114) and a
 Cosmogrid Fellowship (FMR). We also acknowledge financial support
 from the Guangzhou Education Bureau and Guangzhou Science and
 Technology Bureau.


\newpage

\begin{thebibliography}{}

\bibitem{aha99a} Aharonian F., Akhperjinian A.G., Barrio J.A., et al.
             1999a, A\&A, 342, 69

\bibitem{aha99b} Aharonian F., Akhperjanian A.G., Barrio J.A. et al. 1999b,
             A\&A, 349, 11
{
\bibitem{alb07} Albert J., et al. (MAGIC collaboration) 2007, ApJ in press
                (astro-ph/0702008)
}


\bibitem{blo96} Bloom S., Marscher A.P. 1996, ApJ, 463, 555

\bibitem{blu84} Blumenthal G.R., Yang L.T., Lin D.N.C. 1984, ApJ, 287, 774

{
\bibitem{buk96} Buckley, J. H.; Akerlof, C. W.; Biller, S., et al. 1996, ApJ 472 L9
}

\bibitem{cao02} Cao X. 2002, ApJ, 570, L13

\bibitem{cao03} Cao X. 2003, ApJ, 599, 147

\bibitem{cat97} Catanese M., Bradbury S.M., Breslin A.C., et al. 1997, ApJ 487,
                L143

\bibitem{cat99} Catanese M., Weekes T.C. 1999, PASP, 111, 1193

\bibitem{che95} Chen X., Taam R.E. 1995, ApJ, 441, 354

\bibitem{dej99} De Jager O.C., Kranich D., Lorentz E., Kestel M. 1999,
                Proc. 26th ICRC (Salt Lake City), 3, 346

{
\bibitem{der92} Dermer C.D., Schlickeiser R., Mastichiadis A. 1992, A\&A 256, L27

\bibitem{der93} Dermer C.D., Schlickeiser R. 1993, ApJ 416, 458

\bibitem{der94} Dermer C.D., Schlickeiser R. 1994, ApJS, 90, 945
}

\bibitem{der97} Dermer C.D., Sturner S.J., Schlickeiser R. 1997, ApJS, 109, 103

\bibitem{dja99} Djannati-Atai A., Piron F., Barrau A., et al. 1999, A\&A, 350, 17

\bibitem{fal02} Falomo R., Kotilainen J.~K., Treves A.\ 2002, ApJL, 569, L35

\bibitem{fan05} Fan J.H. 2005, A\&A, 436, 799

\bibitem{fan99} Fan J.H., Lin R.G. 1999, ApJS, 121, 131

\bibitem{ghi98} Ghisellini G. 1998, in: The Active X-ray Sky: Results
                from BeppoSAX and RXTE, ed. by L. Scarsi et al.,
                Amsterdam, 397

\bibitem{hay98} Hayashida N., Hirasawa H., Ishikawa F., et al. 1998, ApJL, 504,
                L71

\bibitem{hon92} Honma F., Matsumoto R., Kato S. 1992, PASJ, 44, 529

\bibitem{kat99} Kataoka J., Mattox J.R., Quinn J., et al. 1999, ApJ, 514, 138

{
\bibitem{kat01} Katarzynski K., Sol. H., Kus A. 2001, A\&A 367, 809
}

\bibitem{kat78} Kato S. 1978, MNRAS, 185, 629

\bibitem{kle93} Kley W., Papaloizou J.C.B., Lin D.N.C. 1993, ApJ, 409, 739

\bibitem{kon99} Konopelko A. 1999, APh, 11, 135

{
\bibitem{kon03} Konopelko A., Mastichiadis A., Kirk, J., et al. 2003, ApJ 597,
                851}

\bibitem{kra99} Kranich D., De Jager O.C., Kestel M., et al. 1999,
                Proc. 26th ICRC (Salt Lake City), 3, 358

\bibitem{kra01} Kranich D., De Jager O.C., Kestel M., et al. 2001,
                Proc. 27th ICRC (Hamburg), 7, 2631

\bibitem{kraw00} Krawczynski H., Coppi P.S., Maccarone T., Aharonian F.A. 2000,
                 A\&A, 353, 97

\bibitem{kraw02} Krawczynski H., Coppi P.S., Aharonian F. 2002, MNRAS 336, 721

\bibitem{lam98} Lamer G., Wagner S.J. 1998, A\&A, 331, L13

\bibitem{lao03} Laor A. 2003, ApJ, 590, 86

\bibitem{lig74} Lightman A.P. 1974, ApJ, 194, 419

\bibitem{lige74} Lightman A.P., Eardley D. 1974, ApJ, 187, L1

\bibitem{liu02} Liu F.K., Wu X-B. 2002, A\&A, 388, L48

{
\bibitem{mac95} Macomb, D. J., Akerlof, C. W., Aller, H. D., et al. 1995, ApJ, 449, L99
}

\bibitem{man96} Manmoto T., Takeuchi M., Mineshige S., et al. 1996, ApJ, 464,
                L135

\bibitem{mas04} Massaro E., Perri M., Giommi P., et al. 2004, A\&A, 422, 103

\bibitem{mas97} Mastichiadis A., Kirk J.G. 1997, A\&A, 320, 19

\bibitem{mil97} Milsom J.A., Taam R.E. 1997, MNRAS, 286, 358

\bibitem{nar98} Narayan R., Mahadevan R., Quataert, E.\ 1998, in: Theory of
                Black Hole Accretion Disks, ed. by M.A. Abramowicz et al., p.148

\bibitem{oso06} Osone S., 2006, APh, 26, 209

\bibitem{pap86} Papaloizou J.C.B., Stanley G.Q.R. 1986, MNRAS, 220, 593

\bibitem{pet00} Petry D., B\"ottcher M., Connaughton V., et al 2000, ApJ, 532,
                742

\bibitem{pia98} Pian E., Vacanti G., Tagliaferri G.,  et al. 1998, ApJ, 492, L17

\bibitem{pir78} Piran T. 1978, ApJ, 221, 652

\bibitem{pri76} Pringle J.E. 1976, MNRAS, 177, 65

\bibitem{pro97} Protheroe R.J. et al. 1997, Proc. 25th ICRC (Durban), 8, 317

\bibitem{qui96} Quinn J., Akerlof C.W., Biller S., et al. 1996, ApJ, 456, L83

\bibitem{rie00} Rieger F.M. Mannheim K. 2000, A\&A, 359, 948

\bibitem{rie03} Rieger F.M. Mannheim K. 2003, A\&A, 397, 121

{
\bibitem{rie07} Rieger F.M., Bosch-Ramon V., Duffy P. 2007, Ap\&SS 309, 119
}

\bibitem{rxte99} RXTE 1999, http://space.mit.edu/XTE/asmlc/ASM.html

\bibitem{sam96} Sambruna R., Maraschi L., Urry C.M. 1996, ApJ, 474, 639

\bibitem{sha76} Shapiro S.L., Lightman A.P., Eardley D.M. 1976, ApJ, 204,
                187

\bibitem{sha73} Shakura N.I., Sunyaev R.A. 1973, A\&A, 24, 337

\bibitem{sil88} Sillanp\"a\"a A., Haarala S., Valtonen M. J., et al.
                1988, ApJ, 325, 628

\bibitem{sti93} Stickel M., Fried J.W., K\"uhr H. 1993, A\&AS, 98, 393

\bibitem{tav98} Tavecchio F., Maraschi L., \& Ghisellini G.\ 1998, ApJ, 509,
                608
{
\bibitem{taka96} Takahashi, T.; Tashiro, M.; Madejski, G. et al. 1996, ApJ, 470, L89
}

\bibitem{tav01} Tavecchio F., Maraschi L., Pian E., et al. 2001, ApJ, 554, 725

\bibitem{vil99} Villata M., Raiteri C. M. 1999, A\&A, 347, 30

\bibitem{wan91} Wandel A., Liang E.P. 1991, ApJ, 380, 84

\bibitem{wanu91} Wandel A., Urry M.C. 1991, ApJ, 367, 78

\bibitem{woo05} Woo, J-H., Urry, M.C., et al. 2005, ApJ, 631, 762

\bibitem{wu97} Wu X.B. 1997, MNRAS, 292, 113

\bibitem{yan97} Yang L.T., Henning T., Lu Y., Wu X.B. 1997, MNRAS, 288, 965

\bibitem{yi99} Yi I. 1999, in: Astrophysical Disks, ed. by J.A. Sellwood
               \& J. Goodman, ASP Conf.Ser. 160, 279



\end{thebibliography}
\end{document}